\documentclass[twocolumn,floatfix,rmp]{revtex4}
\usepackage{graphicx}
\usepackage{amsmath}
\usepackage{bm}
\bibpunct{}{}{,}{s}{}{\textsuperscript{,}}
\begin{document}
\title{\Huge\bf\sc Quantum Point Contacts}
\author{\large\bf The quantization of ballistic electron transport
through a constriction\\
demonstrates that conduction is transmission.\bigskip\\
{\small\rm Published in abbreviated form in {\sc Physics Today}, July 1996, page 22.}\bigskip}
\affiliation{\large Henk van Houten \& Carlo Beenakker}

\thanks{{\sc Henk van Houten} {\em heads a department at the Philips Research
Laboratories in Eindhoven, and is a Professor of Physics at the
University of Leiden (The Netherlands).} {\sc Carlo Beenakker} {\em is a
Professor of Physics at the University of Leiden.}}
\maketitle

\noindent
Punctuated equilibrium, the notion that evolution in nature is stepwise
rather than continuous, sometimes applies to evolution in science as
well. It happens that the seed of a scientific breakthrough slumbers
for a decade or even longer, without generating much interest. The seed
can be a theoretical concept without clear predictions to test
experimentally, or an intriguing but confusing experiment without a
lucid interpretation. When the seed finally germinates, an entire field
of science can reach maturity in a few years.

In hindsight, this is what happened ten years ago, when the authors
(newly hired PhD's at Philips Research in Eindhoven) ventured into the
field of quantum ballistic transport. Together with Bart van Wees, then
a graduate student at Delft University of Technology, we were
confronted with some pretty vague challenges. On the experimental side,
there was the search for a quantum-size effect on the conductance,
which would reveal in a clear-cut way the one-dimensional density of
states of electrons confined to a narrow wire. Experiments on narrow
silicon transistors (at Yale University and AT\&T Bell Labs., Holmdel)
had come close, but suffered from irregularities due to disorder.
(These irregularities would become known as ``universal conductance
fluc\-tua\-tions'', see {\small {\sc Physics Today}}, December 1988, page
36.) We anticipated that the electron motion should be ballistic, {\em
i.e.} without scattering by impurities. Moty Heiblum (IBM, Yorktown
Heights) had demonstrated ballistic transport of hot electrons, high
above the Fermi level. For a quantum-size effect one needs ballistic
motion at the Fermi energy. Our colleague Thomas Foxon from Philips
Research in Redhill (UK) could provide us with heterojunctions of GaAs
and \mbox{AlGaAs}, containing at the interface a thin layer of highly
mobile electrons. Such a ``two-dimensional electron gas'' seemed an
ideal system for ballistic transport.

On the theoretical side, there was the debate whether a wire without
impurities could have any resistance at all.\cite{Sto88} Ultimately,
the question was: ``What is measured when you measure a resistance?''
The conventional point of view (held in the classical Drude-Sommerfeld
or the quantum mechanical Kubo theories) is that conduction is the flow
of current in response to an electric field.  An alternative point of
view was put forward in 1957 by Rolf Landauer (IBM, Yorktown Heights),
who proposed that ``conduction is transmission''.\cite{Lan57}
Landauer's formula, a relationship between conductance and transmission
probability, had evolved into two versions. One gave infinite
conductance (= zero resistance) in the absence of impurity scattering,
while the other gave a finite answer.  Although the origin of the
difference between the two versions was understood by at least one of
the theorists involved in the debate,\cite{Imr86} the experimental
implications remained unclear.

Looking back ten years later, we find that the seed planted by Landauer
in the fifties has developed into a sophisticated theory, at the basis
of the entire field of quantum ballistic transport. The breakthrough
can be traced back to experiments on an elementary conductor: a point
contact. In this article we present a brief account of these
developments. For a more comprehensive and detailed discussion, we
direct the reader to the reviews in the bibliography.
\bigskip\\
{\large\bf Quantized conductance}
\medskip\\
The history of ballistic transport goes back to 1965, when Yuri Sharvin
(Moscow) used a pair of point contacts to inject and detect a beam of
electrons in a single-crystalline metal.\cite{Sha65} In such
experiments the quantum mechanical wave character of the electrons does
not play an essential role, because the Fermi wave length
($\lambda_{\rm F}\approx 0.5\,{\rm nm}$) is much smaller than the
opening of the point contact.  The two-dimensional (2D) electron gas in
a GaAs--AlGaAs heterojunction has a Fermi wave length which is a
hundred times larger than in a metal. This makes it possible to study a
constriction with an opening comparable to the wave length (and much
smaller than the mean free path for impurity scattering). Such a
constriction is called a {\em quantum} point contact.

\begin{figure}
\centerline{\includegraphics[width=8cm]{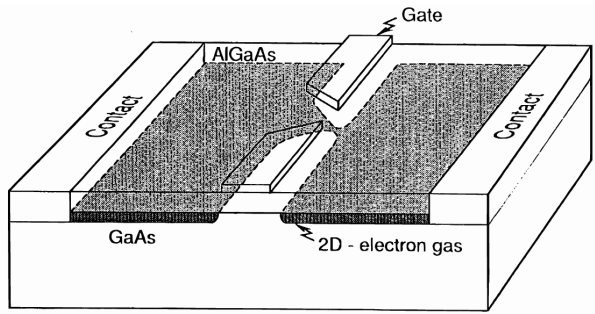}}
\caption{
Schematic cross-sectional view of a quantum point contact, defined in a
high-mobility 2D electron gas at the interface of a GaAs--AlGaAs
heterojunction. The point contact is formed when a negative voltage is
applied to the gate electrodes on top of the AlGaAs layer. Transport
measurements are made by employing contacts to the 2D electron gas at
either side of the constriction.
}\label{fig_device}
\end{figure}

\begin{figure}
\centerline{\includegraphics[width=8cm]{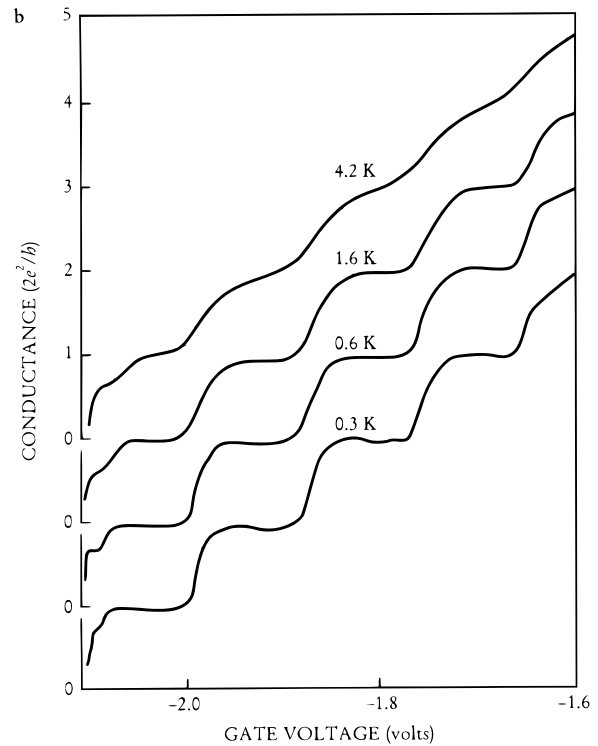}}
\caption{
Conductance quantization of a quantum point contact in units of
$2e^{2}/h$. As the gate voltage defining the constriction is made less
negative, the width of the point contact increases continuously, but
the number of propagating modes at the Fermi level increases stepwise.
The resulting conductance steps are smeared out when the thermal energy
becomes comparable to the energy separation of the modes. (Adapted from
ref.\ \onlinecite{Wee88}.)
}\label{fig_QPC}
\end{figure}

In a metal a point contact is fabricated simply by pressing two wedge-
or needle-shaped pieces of material together. A quantum point contact
requires a more complicated strategy, since the 2D electron gas is
confined at the GaAs--AlGaAs interface in the interior of the
heterojunction. A point contact of adjustable width can be created in
this system using the split-gate technique developed in the groups of
Michael Pepper (Cambridge) and Daniel Tsui (Princeton).\cite{Tho86} The
gate is a negatively charged electrode on top of the heterojunction,
which depletes the electron gas beneath it. (See figure
\ref{fig_device}.) In 1988, the Delft-Philips and Cambridge groups
reported the discovery of a sequence of steps in the conductance of a
constriction in a 2D electron gas, as its width $W$ was varied by means
of the voltage on the gate.\cite{Wee88,Wha88} (See {\sc Physics
Today}, November 1988, page 21.) As shown in figure \ref{fig_QPC}, the
steps are near integer multiples of $2e^{2}/h\approx 1/13\,{\rm
k}\Omega$ (after correction for a small gate-voltage independent series
resistance).

An elementary explanation of the quantization views the constriction as
an electron wave guide, through which a small integer number $N\approx
2W/\lambda_{\rm F}$ of transverse modes can propagate at the Fermi
level. The wide regions at opposite sides of the constriction are {\em
reservoirs\/} of electrons in local equilibrium. A voltage difference
$V$ between the reservoirs induces a current $I$ through the
constriction, equally distributed among the $N$ modes. This
equipartition rule is not immediately obvious, because electrons at the
Fermi level in each mode have different group velocities $v_{n}$.
However, the difference in group velocity is canceled by the difference
in density of states $\rho_{n}=1/hv_{n}$. As a result, each mode
carries the same current $I_{n}=Ve^{2}\rho_{n}v_{n}=Ve^{2}/h$.  Summing
over all modes in the wave guide, one obtains the conductance
$G=I/V=Ne^{2}/h$.  The experimental step size is twice $e^{2}/h$
because spin-up and spin-down modes are degenerate.

The electron wave guide has a non-zero resistance even though there are
no impurities, because of the reflections occurring when a small number
of propagating modes in the wave guide is matched to a larger number of
modes in the reservoirs. A thorough understanding of this mode-matching
problem is now available, thanks to the efforts of many
investigators.\cite{Eer}

The quantized conductance of a point contact provides firm experimental
support for the Landauer formula,
\[
G=\frac{2e^{2}}{h}\sum_{n}t_{n},
\]
for the conductance of a disordered metal between two electron
reservoirs. The numbers $t_{n}$ between 0 and 1 are the eigenvalues of
the product ${\bf tt}^{\dagger}$ of the transmission matrix ${\bf t}$
and its Hermitian conjugate. For an ``ideal'' quantum point contact $N$
eigenvalues are equal to 1 and all others are equal to 0. Deviations
from exact quantization in a realistic geometry are about 1\%. This can
be contrasted with the quantization of the Hall conductance in strong
magnetic fields, where an accuracy better than 1 part in $10^{7}$ is
obtained routinely.\cite{Cag90} One reason why a similar accuracy can
not be achieved in zero magnetic field is the series resistance from
the wide regions, whose magnitude can not be determined precisely.
Another source of excess resistance is backscattering at the entrance
and exit of the constriction, due to the abrupt widening of the
geometry. A magnetic field suppresses this backscattering, improving
the accuracy of the quantization.

Suppression of backscattering by a magnetic field is the basis of the
theory of the quantum Hall effect developed by Marcus B\"{u}ttiker
(IBM, Yorktown Heights).\cite{But88} B\"{u}ttiker's theory uses a
multi-reservoir generalization of the two-reservoir Landauer formula.
The propagating modes in the quantum Hall effect are the magnetic
Landau levels interacting with the edge of the sample. (Classically,
these magnetic edge states correspond to the skipping orbits discussed
later.) There is a smooth crossover from zero-field conductance
quantization to quantum Hall effect, corresponding to the smooth
crossover from zero-field wave guide modes to magnetic edge states.
\bigskip\\
{\large\bf When 1 mode = 1 atom}
\medskip\\
Since the conductance quantum $e^2/h$ contains only constants of
nature, the conductance quantization might be expected to occur in
metals as well as in semiconductors.  A quantum point contact in a
semiconductor is a {\em mesoscopic\/} object, on a scale intermediate
between the macroscopic world of classical mechanics and the
microscopic world of atoms and molecules. This separation of length
scales exists because of the large Fermi wave length in a
semiconductor. In a metal, on the contrary, the Fermi wave length is of
the same order of magnitude as the atomic separation. A quantum point
contact in a metal is therefore necessarily of atomic dimensions.

If the initial contact between two pieces of metal is formed by a
single atom, the conductance will be of the order of $2e^{2}/h$. This
was first observed in 1987 by James K. Gimzewski and R. M\"{o}ller
(IBM, Z\"{u}rich),\cite{Gim87} in experiments where the Ir tip of a
scanning tunneling microscope was pressed onto a Ag surface. Upon
making contact, the conductance jumped from an exponentially small
value to a value of $1/16\,{\rm k}\Omega$. Later work, using
mechanically more stable devices, showed that further jumps of order
$2e^{2}/h$ in the conductance occur as the contact area is increased.

\begin{figure}
\centerline{\includegraphics[width=8cm]{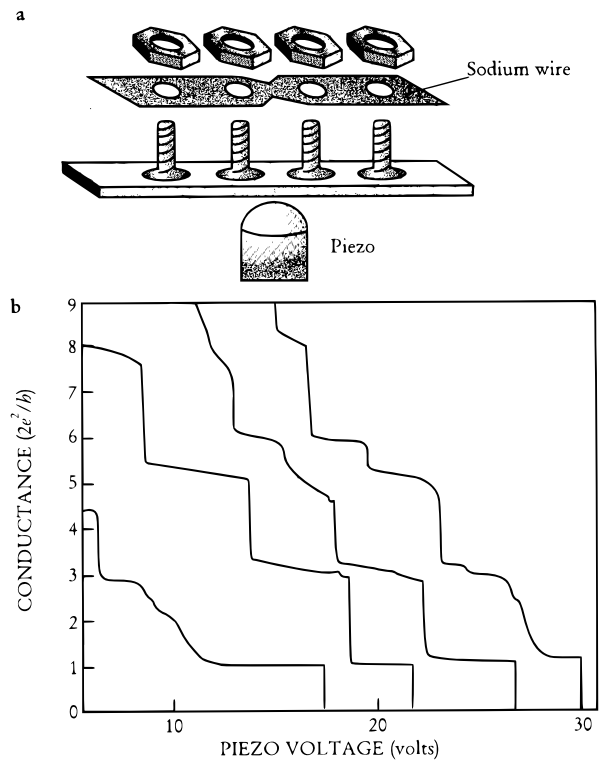}}
\caption{
Quantized steps in the conductance of a Na point contact. A clean point
contact is made at 4.2~K by breaking the wire at a notch cut across a
thin wire, whereafter the two parts are brought into contact
mechanically. The width of the point contact is adjusted by increasing
the force of contact through a piezo element.  Electrical measurements
are made using four miniature brass bolts connected to the wire.
The experiment is not fully reproducible, because of different
atomic rearrangements in the contact region. (Adapted from
ref.\ \onlinecite{Kra95}.)
}\label{fig_Na}
\end{figure}

Figure \ref{fig_Na} shows experimental data for a Na point contact
measured by Martijn Krans and collaborators from the Kamerlingh Onnes
Laboratory in Leiden.\cite{Kra95} An adjustable contact of atomic
dimensions, with a high mechanical stability, is made by bolting a
notched wire of sodium onto a flexible substrate. As the substrate is
bent, the wire breaks at the notch. The contact area can be controlled
down to the atomic scale, simply by bending the substrate more or less.
A statistical analysis of a large number of samples shows that, as the
contact area is increased, steps in the conductance appear near 1, 3,
5, and 6 times $2e^{2}/h$. (Figure \ref{fig_Na} shows the conductance
steps for representative single measurements.) The absence of steps at
2 and 4 times $2e^{2}/h$ is significant, and has a neat explanation: In
a cylindrically symmetric potential the second and third transverse
mode are degenerate, as are the fourth and fifth mode, while the first
and sixth mode are non-degenerate.

\begin{figure}
\centerline{\includegraphics[width=8cm]{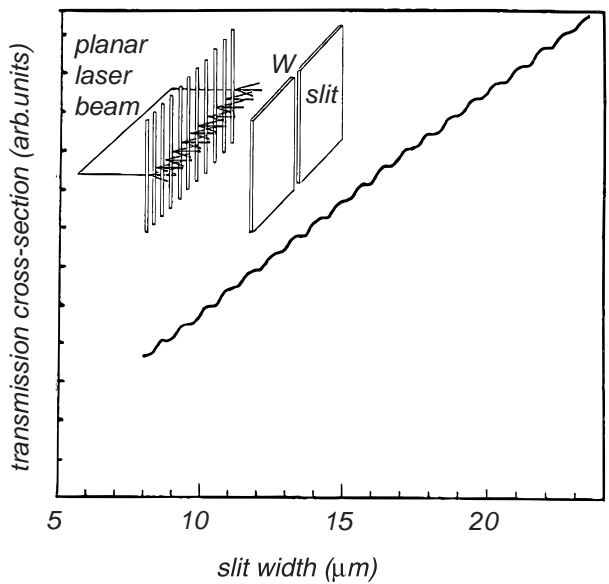}}
\caption{
Equidistant steps in the optical transmission cross-section of a slit
of adjustable width. A 2D isotropic illumination is obtained by shining
a beam from a diode laser ($\lambda=1.55\,\mu{\rm m}$, polarization
parallel to the slit) onto a diffusor consisting of a random array of
parallel glass fibers.  (Adapted from ref.\ \onlinecite{Mon91}.)
}\label{fig_light}
\end{figure}

The energy separation of transverse modes in a point contact of atomic
dimensions is so large that the conductance steps are visible at room
temperature. Nicol\'{a}s Garc\'{\i}a and his group at the Autonomous
University of Madrid have made use of this property to develop a
classroom experiment of quantized conductance. (See {\sc Physics
Today}, February 1996, page 9.)
\bigskip\\
{\large\bf Photons and Cooper pairs}
\medskip\\
The interpretation of conduction as transmission of electrons at the
Fermi level suggests an analogy with the transmission of monochromatic
light.\cite{Hou90} The analogue of the conductance is the transmission
cross-section $\sigma$, defined as the transmitted power divided by the
incident flux. Figure \ref{fig_light} shows the transmission cross-section of a slit
of variable width, measured by Edwin Montie and collaborators from
Philips.\cite{Mon91} Steps of equal height occur whenever the slit
width $W$ equals half the wave length $\lambda=1.55\,\mu{\rm m}$ of the
light. Because $\sigma$ equals $W$ for large slit widths, the step
height is also equal to $\lambda/2$. Two-dimensional isotropic
illumination was achieved by passing the light through a random array
of glass fibres parallel to the slit. The isotropy of the illumination
mimics the reservoirs in the electronic case, and is crucial for the
effect. The two-dimensionality is not essential, but was chosen because
a diaphragm of variable area of the order of $\lambda^{2}$ is difficult
to fabricate. (For a diaphragm, the steps in $\sigma$ are
$\lambda^{2}/2\pi$.)

It is remarkable that this optical phenomenon, with its distinctly
nineteenth century flavour, was not noticed prior to the discovery of
its electronic counterpart. There is an interesting parallel in the
history of the discovery of the two phenomena. In the electronic case,
the Landauer formula was already known before the quantized conductance
of a point contact was discovered. Yoseph Imry (Weizmann Institute,
Israel) had made the connection with Sharvin's work on point
contacts.\cite{Imr86} The reason that the conductance quantization came
as a surprise, was that the relation $\sum_{n}t_{n}=N$ for ballistic
transport was regarded as an order-of-magnitude estimate. To have
quantization, the relative error in this estimate must be smaller than
$1/N$, which is not obvious. The equivalent of the Landauer formula for
the transmission cross-section has long been familiar in
optics,\cite{Sny73} but also in this field it was not noticed that
$\sum_{n}t_{n}=N$ holds with better than $1/N$ relative accuracy.

One can speak of the optical analogue as a quantum point contact for
photons. Can the analogue be extended towards a quantum point contact
for Cooper pairs? The answer is ``Yes'': The maximal supercurrent
through a narrow and short, impurity-free constriction in a
superconductor is an integer multiple of $e\Delta/\hbar$, with $\Delta$
the energy gap of the bulk superconductor.\cite{Bee91} A
superconducting quantum point contact has been realized by Hideaki
Takayanagi and collaborators (NTT, Japan),\cite{Tak95} but the
superconducting analogue of the quantized conductance remains to be
observed experimentally.
\bigskip\\
{\large\bf Thermal analogues}
\medskip\\
The conductance is the coefficient of proportionality between
current and voltage. The additional presence of a small
temperature difference $\delta T$ across the point contact gives rise
to a matrix of coefficients:
\[
\left(
\begin{array}{c}
\mbox{electrical current}\\
\mbox{heat current}
\end{array}
\right)
=
\left(
\begin{array}{cc}
G&L\\
L'&K
\end{array}
\right)
\cdot
\left(
\begin{array}{c}
- V\\
\delta T
\end{array}
\right).
\]
The thermal conductance $K$ relates heat current to temperature
difference. The thermo-electric cross-phenomena are described by
coefficients $L$ and $L'$. As first deduced by Lord Kelvin,
time-reversal symmetry requires that $L'=-LT$ (at a temperature $T$).

The two new transport coefficients $K$ and $L$ can be expressed in
terms of the transmission probabilities, just like the electrical
conductance $G$. (See sidebar.) Approximately, $K\propto t$ and
$L\propto dt/dE_{\rm F}$, where $t=\sum_{n}t_{n}$ is the total
transmission probability at the Fermi energy $E_{\rm F}$.  (The
proportionality of $K$ to $t$, and hence to $G$, is the Wiedemann-Franz
law of solid-state physics.) The stepwise energy dependence of the
transmission probability through a quantum point contact implies two
types of quantum-size effects: steps in $K$ and peaks in $L$. Both
effects have been observed by Laurens Molenkamp and collaborators from
Philips.\cite{Hou92}

The thermal conductance $K$ of a quantum point contact exhibits steps
when the gate voltage is varied, aligned with the steps in the
electrical conductance. Each step signals the appearance of a new mode
at the Fermi level which can propagate through the constriction. A step
in the transmission probability leads to a peak in the thermo-electric
coefficient $L$. Pavel St\u{r}eda\cite{Str89} (Prague) has calculated
that, at zero temperature, the height of the peaks in $L$ is
approximately $k/e$ times the conductance quantum $e^{2}/h$. The unit
$k/e\approx 50\,\mu{\rm V}/{\rm K}$ is the entropy production per
Coulomb of charge transferred through the point contact, or $1/e$ times
the entropy carried by a single conduction electron, which is on the
order of Boltzmann's constant $k$.

\begin{figure}
\centerline{\includegraphics[width=8cm]{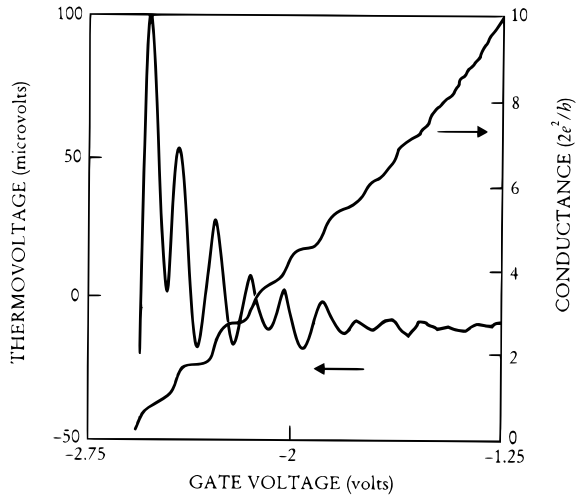}}
\caption{
Thermopower oscillations in a quantum point contact. The peaks in the
thermovoltage (proportional to the thermopower) coincide with the steps
in the conductance. (Adapted from ref.\ \onlinecite{Mol90a}.)
}\label{fig_thermo}
\end{figure}

In figure \ref{fig_thermo} we show measurements of the thermopower
$S=-L/G$ of a quantum point contact.\cite{Mol90a} (The thermopower is
proportional to the voltage produced by a temperature difference for
zero electrical current.) The coincidence of peaks in the thermopower
with steps in the conductance (measured for the same point contact) is
clearly visible. Joule heating was used to create a temperature
difference across the point contact in this work. Local heating by
means of a focused beam of far-infrared radiation has been used in a
more recent experiment.\cite{Wys95} \bigskip\\
{\large\bf Shot noise}
\medskip\\
The electrical current through a point contact is not constant in time,
but fluctuates. The conductance determines only the time-averaged
current. The noise power $P=2\int dt\,\langle\delta I(0)\delta
I(t)\rangle\cos\omega t$ at frequency $\omega$ is the Fourier transform
of the correlator of the time-dependent fluctuations $\delta I(t)$ in
the current at a given voltage $V$ and temperature $T$. One
distinguishes equilibrium thermal noise ($V=0$, $T\neq 0$) and
non-equilibrium shot noise ($V\neq 0$, $T=0$). Both types of noise have
a white power spectrum ({\em i.e.} the noise power does not depend on
frequency over a very wide frequency range). Thermal noise is directly
related to the conductance through the fluctuation-dissipation theorem
($P_{\rm thermal}=4kTG$).  Therefore, the thermal noise of a quantum
point contact does not give any new information.

Shot noise is more interesting, because it contains information on the
temporal correlation of the electrons which is not contained in the
conductance. Maximal shot noise ($P_{\rm max}=2eI$) is observed when
the stream of electrons is fully uncorrelated. A typical example is a
tunnel diode. Correlations reduce $P$ below $P_{\rm max}$. One source of
correlations, operative even for non-interacting electrons, is the
Pauli principle, which forbids multiple occupation of the same
single-particle state. A typical example is a ballistic point contact
in a metal, where $P=0$ because the stream of electrons is completely
correlated by the Pauli principle in the absence of impurity
scattering.

\begin{figure}
\centerline{\includegraphics[width=8cm]{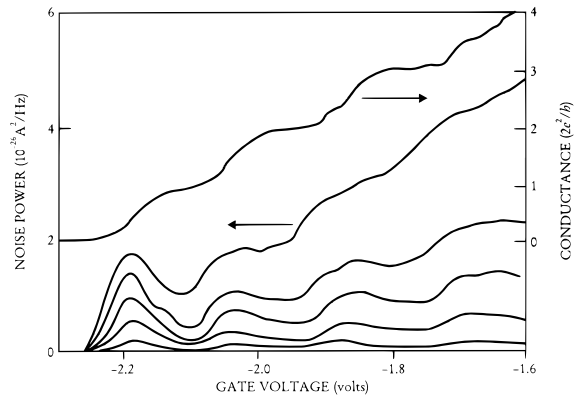}}
\caption{
Periodic suppression of the shot-noise power of a quantum point
contact, measured with an applied voltage of 0.5, 1, 1.5, 2, and 3 mV.
(Adapted from ref.\ \onlinecite{Rez95}.)
}\label{fig_shot}
\end{figure}

A quantum point contact in a 2D electron gas has a different behavior.
Using a Landauer-type formula (see sidebar), Gordey Lesovik (Moscow)
has predicted peaks in the shot noise at the steps in the
conductance.\cite{Les89} The peak height $P_{\rm peak}=eI$ is half the
maximal value for uncorrelated electrons. The shot noise vanishes in
between the steps. Michael Reznikov and collaborators from the Weizmann
Institute in Israel have recently presented a convincing demonstration
of this quantum-size effect in the shot noise.\cite{Rez95} (See figure
\ref{fig_shot}.) By going to microwave frequencies (8--18~GHz) they
could avoid the ubiquituous ``1/f noise'' at lower frequencies.
\bigskip\\
{\large\bf Solid-state electron optics}
\medskip\\
The effects discussed so far refer to properties of the quantum point
contact itself. A wealth of new phenomena has been discovered using a
quantum point contact as a spatially coherent point source and
detector, and specially formed electrodes as mirror, prism, or lens.

\begin{figure}
\centerline{\includegraphics[width=8cm]{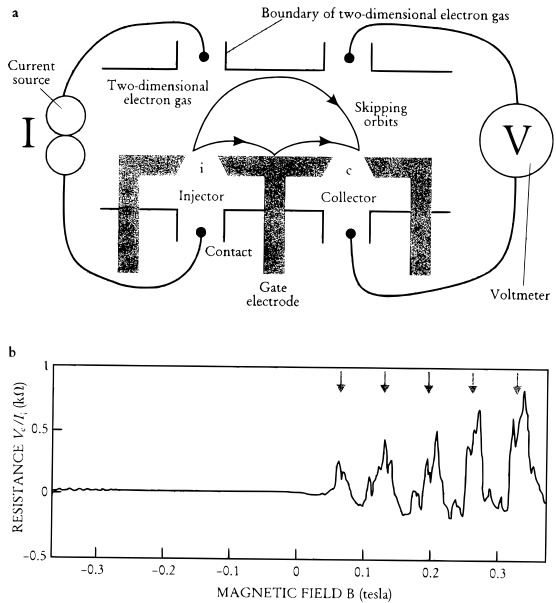}}
\caption{
Magnetic focusing in a 2D electron gas at $50\,{\rm mK}$. The top panel
shows the experimental arrangement. Electrons injected through one
point contact (i) follow skipping orbits over a distance of
$3.0\,\mu{\rm m}$ to a second point contact (c). The arrows indicate
the positions of the focusing peaks expected when the point contact
separation is a multiple of the cyclotron diameter. The fine-structure
on the peaks is due to quantum interference. (Adapted from
ref.\ \onlinecite{Hou88}.)
}\label{fig_cef}
\end{figure}

The basic experiment,\cite{Hou88} {\em coherent electron focusing}, is
shown in figure\ \ref{fig_cef}. A point contact injects electrons with
the Fermi momentum $p_{\rm F}$ into the 2D electron gas, in the
presence of a perpendicular magnetic field $B$. The electrons follow a
``skipping orbit'' along the boundary, consisting of circular arcs of
cyclotron diameter $d_{\rm c}=2p_{\rm F}/eB$. Some of the electrons are
collected at a second point contact, at a separation $L$ from the
first. The voltage measured at the collector is proportional to the
transmission probability between the two point contacts. V. S.
Tso\u{\i} (Moscow) first used this focusing technique in a
metal.\cite{Tso74} The magnetic field acts as a lens, bringing the
divergent trajectories at the injector together at the collector. The
collector is at a focal point of the lens when $L$ is a multiple of
$d_{\rm c}$, hence when $B$ is a multiple of $2p_{\rm F}/eL$ (arrows in
figure \ref{fig_cef}). For reverse magnetic fields the injected
electrons are deflected away from the collector, so that no signal is
generated. The observation of peaks at the expected positions
demonstrates that a quantum point contact acts as a monochromatic point
source of ballistic electrons, and that the reflections at the boundary
of the 2D electron gas are specular. The fine-structure on the focusing
peaks is due to quantum interference of trajectories between the two
point contacts. Such fine-structure does not appear in metals. It
demonstrates that the quantum point contact is a spatially coherent
source and that the phase coherence is maintained over a distance of
several microns to the collector.

Magnetic focusing has been used by several groups to obtain information
on the dynamics and scattering of quasiparticles in the 2D electron
gas. An intriguing application in the regime of the fractional quantum
Hall effect is the focusing of composite fermions,\cite{Gol94} which
can be thought of as electrons bound to an even number of flux quanta.
In the regime of the integer quantum Hall effect, the geometry of
figure \ref{fig_cef} has been used to selectively populate and detect
the magnetic edge states mentioned earlier.\cite{Wee88} The observation of
plateaus in the Hall conductance at anomalously quantized values provides
support for the edge-state theory of the quantum Hall effect.

Electrostatic focusing, by means of the electric field produced by a
lens-shaped electrode, provides an alternative technique to focus the
beam of electrons injected by a point contact.\cite{Siv90} Instead of
focusing the beam, one can also deflect it --- either by means of a
magnetic field,\cite{Mol90b} or by means of a prism-shaped
electrode.\cite{Spe90} The building blocks of electron optics in the
solid state have by now all been realized.
\bigskip\\
{\large\bf Ultimate confinement}
\medskip\\
A quantum point contact which is nearly pinched off (so that its
conductance is less than $2e^{2}/h$) is a tunnel barrier of adjustable
height for electrons near the Fermi level. This property has been used
to inject and detect electrons in a small confined region of a 2D
electron gas, called a {\em quantum dot}. A quantum dot coupled to the
outside by a pair of quantum point contacts has provided an ideal model
system for the investigation of the effects of Coulomb repulsion on
resonant tunneling. (See {\sc Physics Today}, January 1993, page
24.)

The zero-dimensional quantum dot forms the logical end to the reduction
of dimensionality of the two-dimensional electron gas. In this article
we have reviewed the role played by the one-dimensional quantum point
contact in the conceptual development started by Landauer four decades
ago. The concept of electrical conductance was conceived in the
nineteenth century, at a time when the electron was not even
discovered. It is amusing that it required the sophisticated
micro-electronics technology of the late twentieth century to
demonstrate experimentally that ``conduction is transmission''.
\vspace{2cm}

\noindent
\rule{8.5cm}{1mm}\medskip\\
{\large\bf Landauer formulas}
\medskip\\
Landauer's original 1957 formula,\cite{Lan57}
\[
G=\frac{2e^{2}}{h}\frac{t}{1-t},
\]
expresses the conductance of a one-dimensional system as the ratio of
transmission and reflection probabilities. As explained by
Imry,\cite{Imr86} this formula gives infinity for unit transmission
because it excludes the finite contact conductance contained
in\cite{Eco81}
\[
G=\frac{2e^{2}}{h}t.
\]
Extension to higher dimensions\cite{Fis81} is achieved by replacing the
transmission probability $t$ by the eigenvalue $t_{n}$ of the
transmission matrix product ${\bf t}{\bf t}^{\dagger}$, and summing
over $n$.

Generalizations of the Landauer formula have been found for a variety
of other transport properties, besides the conductance.\cite{credits}
At zero temperature, these expressions are of the form:
\[
\mbox{transport property}=A_{0}\sum_{n}a(t_{n}),
\]
\begin{itemize}
\item conductance $G$: $A_{0}=2e^{2}/h$, $a(t)=t$.
\item shot-noise power $P$: $A_{0}=4e^{3}V/h$, $a(t)=t(1-t)$.
\item conductance $G_{\rm NS}$ of a normal-metal -- superconductor
junction: $A_{0}=4e^{2}/h$, $a(t)=t^{2}(2-t)^{-2}$.
\item supercurrent $I$ through a Josephson junction with phase
difference $\phi$:
$A_{0}=e\Delta/\hbar$,
$a(t)=\frac{1}{2}t\sin\phi(1-t\sin^{2}\phi/2)^{-1/2}$.
\end{itemize}

The expressions for the thermo-electric coefficients involve an
integration over energies around the Fermi energy $E_{\rm F}$, weighted
by the derivative $f'=df/dE$ of the Fermi-Dirac distribution
function at temperature $T$:
\[
\mbox{transport property}=-A_{0}\int dE\,(E-E_{\rm
F})^{p}f'\sum_{n}t_{n},
\]
\begin{itemize}
\item electrical conductance $G$: $p=0$, $A_{0}=2e^{2}/h$.
\item thermo-electric coefficient $L$: $p=1$, $A_{0}=2e/hT$.
\item thermal conductance $K$: $p=2$, $A_{0}=-2/hT$.
\end{itemize}
If the energy-dependence of the transmission eigenvalues is small on
the scale of the thermal energy $kT$, one has approximately
$K=-L_{0}TG$ (the Wiedemann-Franz law) and $L=eL_{0}TdG/dE_{\rm
F}$, with $L_{0}=\pi^{2}k^{2}/3e^{2}$ the Lorentz number.

Each Landauer formula predicts a specific quantum-size effect in a
ballistic constriction, for which $t_{n}$ equals 0 or 1. The effect is
a step-function dependence on the width of the constriction in the case
of $G$, $G_{\rm NS}$, $I(\phi)$, $K$, and an oscillatory dependence in
the case of $P$, $L$.\medskip\\
\rule{8.5cm}{1mm}

\end{document}